\documentclass[twocolumn,tighten,times]{aastex631}
\usepackage{graphicx}

\def\gtsim {>\kern-1.2em\lower1.1ex\hbox{$\sim$}~}   
\def\ltsim {<\kern-1.2em\lower1.1ex\hbox{$\sim$}~}   

\received{July 31, 2023}
\revised{September 3, 2023}
\accepted{September 4, 2023}

\shorttitle{On the $\alpha$/Fe bimodality of the M31 disks}
\shortauthors{Kobayashi, Bhattacharya et al.}

\begin{document}

\title{On the $\alpha$/Fe bimodality of the M31 disks}

\author[0000-0002-4343-0487]{Chiaki Kobayashi}
\affiliation{Centre for Astrophysics Research, Department of Physics, Astronomy and Mathematics, University of Hertfordshire, Hatfield, AL10 9AB, UK}
\email{c.kobayashi@herts.ac.uk}

\author[0000-0003-4594-6943]{Souradeep Bhattacharya}
\affiliation{Inter University Centre for Astronomy and Astrophysics, Ganeshkhind, Post Bag 4, Pune 411007, India}
\author[0000-0001-7214-3009]{Magda Arnaboldi}
\affiliation{European Southern Observatory, Karl-Schwarzschild-Str. 2, 85748 Garching, Germany }
\author[0000-0003-3333-0033]{Ortwin Gerhard}
\affiliation{Max-Planck-Institut für extraterrestrische Physik, Giessenbachstraße, 85748 Garching, Germany}

\begin{abstract}
An outstanding question is whether the $\alpha$/Fe bimodality exists in disk galaxies other than in the Milky Way.
Here we present a bimodality using our state-of-the-art galactic chemical evolution models that can explain various observations in the Andromeda Galaxy (M31) disks, namely, elemental abundances both of planetary nebulae, and of red-giant branch stars recently observed with the James Webb Space Telescope.
We find that in M31 a high-$\alpha$ thicker-disk population out to 30 kpc formed by more intense initial star burst than in the Milky Way.
We also find a young low-$\alpha$ thin disk within 14 kpc, which is formed by a secondary star formation M31 underwent about 2--4.5 Gyr ago, probably triggered by a wet merger.
In the outer disk, however, the planetary nebula observations indicate a slightly higher-$\alpha$ young ($\sim2.5$ Gyr) population at a given metallicity, possibly formed by secondary star formation from almost pristine gas.
Therefore, an $\alpha$/Fe bimodality is seen in the inner disk ($\ltsim14$ kpc), while only a slight $\alpha$/Fe offset of the young population is seen in the outer disk ($\gtsim18$ kpc).
The appearance of the $\alpha$/Fe bimodality depends on the merging history at various galactocentric radii, and wide-field multi-object spectroscopy is required for unveiling the history of M31.
\end{abstract}


\section{Introduction}

The [$\alpha$/Fe]--[Fe/H] relation in a galaxy is one of the most important diagrams in galactic chemical evolution (GCE) as it can tell the formation history of the galaxy.
$\alpha$ elements (O, Ne, Mg, Si, S, Ar, and Ca) are mainly produced from core-collapse supernovae on a short timescale (3--20 Myrs), while the majority of Fe-peak elements are produced from Type Ia supernovae (SNe Ia) on a longer timescale (0.1--20 Gyrs). 
Therefore, the [$\alpha$/Fe] ratios show roughly a constant value at low metallicities (e.g., [Fe/H]) until the SN Ia enrichment becomes significant.
By comparing cutting-edge observations of elemental abundances with GCE models, it is possible to constrain gas accretion, galaxy mergers, and resultant star burst events, for Andromeda Galaxy (M31) more robustly than in previous modelling \citep[e.g.,][]{renda05,yin09,marcon10}.

In the Milky Way (MW), high-resolution ($R\gtsim40,000$) observations of nearby stars and 3D and non-local thermodynamic equilibrium analysis \citep{zhao16,ama19b} can provide the most accurate measurements of elements, which showed a plateau of $[\alpha$/Fe] and its decreasing trend from [Fe/H] $\sim-1$ in the solar neighborhood.
Much larger samples of medium-resolution ($R\sim20,000$) spectroscopic surveys (e.g., APOGEE, HERMES-GALAH) confirmed a bimodal distribution along the [$\alpha$/Fe]--[Fe/H] relation \citep{hayden15}, as indicated in earlier works of high-resolution observations \citep{fuhrmann98,ben04} and a chemodynamical simulation \citep{kob11mw}.

In the [$\alpha$/Fe] ratios of the MW stars, a gap is seen at [Fe/H] $\sim\,-1$, often attributed to thin and thick disk populations, and the origin of this bimodality has been debated using various theoretical models. While `one-zone' GCE models \citep[e.g.,][]{chi97,spitoni19} can explore model parameters that can explain the average trend of each component, chemodynamical simulations \citep[e.g.,][]{kob11mw,brook12,grand18,mackereth18,clarke19,buck20,vin20} can predict the number of stars in the [$\alpha$/Fe]--[Fe/H] diagram.
The MW surveys \citep{hayden15,Guiglion23} also showed that 
the $\alpha$/Fe bimodality 
depends on the location within the Galaxy (galactocentric radius and the height from disk plane). Such dependence is also seen in chemodynamical simulations  
(Fig.\,1 of \citealt{kob16iau}; Fig.\,3 of \citealt{vin20}).
In these cosmological `zoom-in' simulations, the key factor for producing the bimodality is the rapid decrease of $\alpha$/Fe ratios due to the Fe production by SNe Ia, and similar bimodality is expected for many disk galaxies \S 2.2 for more details).

The question we would like to address is whether a similar $\alpha$/Fe dichotomy exists in all disk galaxies or not. 
The closest analogue ($\sim$776~kpc; \citealt{Savino22}) where we can resolve stars to measure elemental abundances is M31.
\citet{vargas14} was the first to do this for only four halo red giant branch (RGB) stars, which is expanded to 70 stars by \citet{escala20} including M31's outer disk.
In \citet{Arnaboldi22}, we showed a bimodality using planetary nebulae (PNe\footnote{A short-lived emission-line nebula phase of intermediate-mass stars (having initial masses of $\sim$0.8--8~M$_{\odot}$), thus covering an age range of $\sim$0.3--10~Gyr. PNe are excellent tracers of light, kinematics and chemistry in galaxies \citep[see review by][]{Kwitter22}.}) but for O/Ar ratios.
Although Ar is an $\alpha$ element, 34\% of Ar comes from SNe Ia \citep{kob20sr}, and thus we can use Ar as proxy for Fe.
In our O/Ar--Ar/H diagram, ``thick disk'' PNe show higher O/Ar than ``thin disk'' PNe at a given metallicity (Ar/H in our case), but this bimodality was seen only at galactocentric radius $R_{\rm GC}\ltsim14$ kpc but not in the outer region ($R_{\rm GC}\gtsim18$).

The James Webb Space Telescope (JWST) broke the observational limits for RGB stars identified with NIRCam photometry and follow-up spectroscopy with NIRSpec \citep{nidever23}. They presented the [$\alpha$/Fe]--[Fe/H] relation in M31, but unfortunately at $R_{\rm GC}\sim18$ kpc, where we did not see the bimodality with PNe, and neither did they with RGB stars (see \S 2.1 for more details).
In this paper, using the same chemical evolution model used in \citet{Arnaboldi22} for PNe (\S 2), we show, for the first time, predictions of the [$\alpha$/Fe]--[Fe/H] relations inner and outer disks (\S 3). \S 4 denotes our conclusions.

\section{Our models}
\subsection{Observational constraints for M31}

M31 is the only other massive disk galaxy where [$\alpha$/Fe] has been measured for resolved individual stars in the nearby universe.
The total mass of M31, at $\sim 1.5 \times 10^{12}$~M$_{\odot}$ \citep[see][and references therein]{Bhattacharya23b}, is almost twice as much as the MW. The total mass and scale radius of M31 disk are also twice those of the MW disk \citep{yin09}. M31 has a larger bulge but a weaker bar than the MW \citep[e.g.,][]{BlanaDiaz18}.
The last major (mass ratio$\sim$1:4) merger epoch is estimated as $\sim$2--4 billion years in M31 \citep{Hammer18,Bh+19b,Bhattacharya23a}, while it is $>9$ billion years in the MW \citep{Belokurov18,Helmi18}.

\begin{figure}
	\includegraphics[width=\columnwidth]{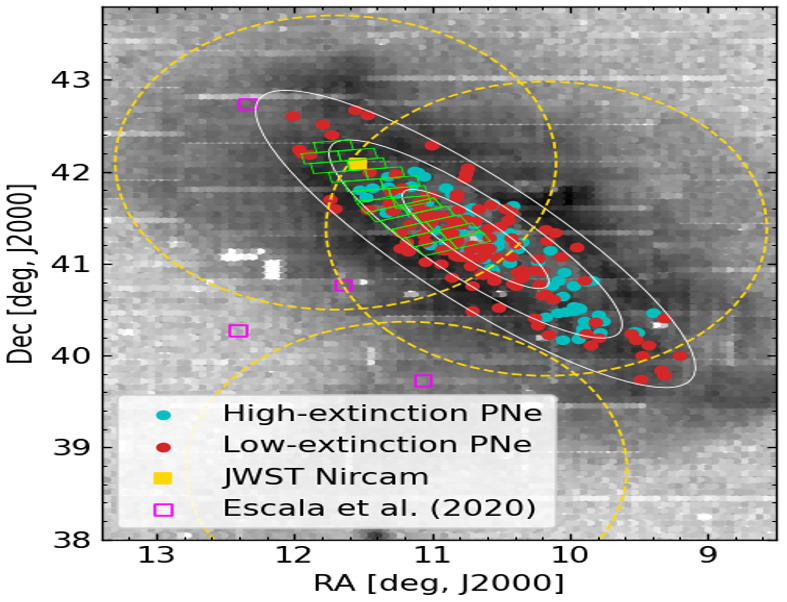}
    \caption{The number density map of RGB stars from PAndAS \citep{mcc18}, binned for visual clarity, is shown in grey. 
    The white ellipses show R$\rm_{GC}=$10,~20,~30~kpc respectively. 
    High- (thin disk) and low-extinction (thick disk) PNe with chemical abundance measurements \citep{Bhattacharya22} are marked in blue and red, respectively. The PHAT survey bricks \citep{dal12} are marked in green. The positions of the Keck/Deimos fields \citep[magenta squares]{escala20}, the three DESI M31 fields, two in the disk, \citep[dashed yellow circles]{dey23}, and the JWST/NIRCam observation \citep[yellow squares]{nidever23} are also marked.}
    \label{fig:spat}
\end{figure}

\citet{vargas14} measured [$\alpha$/Fe] ratios of four halo red giant branch (RGB) stars using deep Keck/Deimos spectroscopy and a grid of synthetic stellar spectra. This was expanded to 21 stars in the Giant Stellar Stream (GSS) substructure by \citet{gilbert19}, and 70 stars by \citet{escala20} for the inner halo, the GSS, and the outer disk; these fields are marked in Figure \ref{fig:spat} (magenta squares). The outer disk field is located at $R_{\rm GC} \sim 31$~kpc where \citet{escala20} measured the mean [Fe/H]= $-0.82 \pm 0.09$ and mean [$\alpha$/Fe]= $0.6^{+0.09}_{-0.1}$ for 18 RGBs.

For integrated lights, using the VIRUS-W integral field spectrograph at the McDonald Observatory 2.7 m telescope, \citet{Saglia18} derived the age, [M/H] and [$\alpha$/Fe] in the M31 bulge and parts of the inner disk within R$\rm_{GC}\sim8$~kpc. They found a near constant [$\alpha$/Fe] = $0.27 \pm 0.08$ for the M31 disk (having age $\sim 9.7 \pm 1.1$~Gyr). A similar value of [$\alpha$/Fe] = $0.274^{0.020}_{-0.025}$ but with age $>12$~Gyr was also measured for the M31 disk within R$\rm_{GC}\sim7$~kpc from APOGEE integrated light spectra \citep{gibson23}. Both these studies thus found an $\alpha$-enhanced relatively old disk in M31 compared to the MW, but it is very difficult to decompose into thin and thick disks from these data.

From the survey of PNe in M31 using narrow- and broad-band imaging from CFHT/Megacam \citep{Bh+19,Bh21} and follow-up multi-object spectroscopy using MMT/Hectospec, the kinematically and chemically distinct thin and thick disks of M31 were identified \citep{Bh+19b,Bhattacharya22}. In particular, \citet{Bhattacharya22} measured O and Ar abundances separately for $\sim$2.5~Gyr old high-extinction PNe that form the ``thin disk'' of M31, and for $>$4.5~Gyr old low-extinction PNe that form the M31 ``thick disk''. 

\begin{figure*}
	\includegraphics[width=\textwidth]{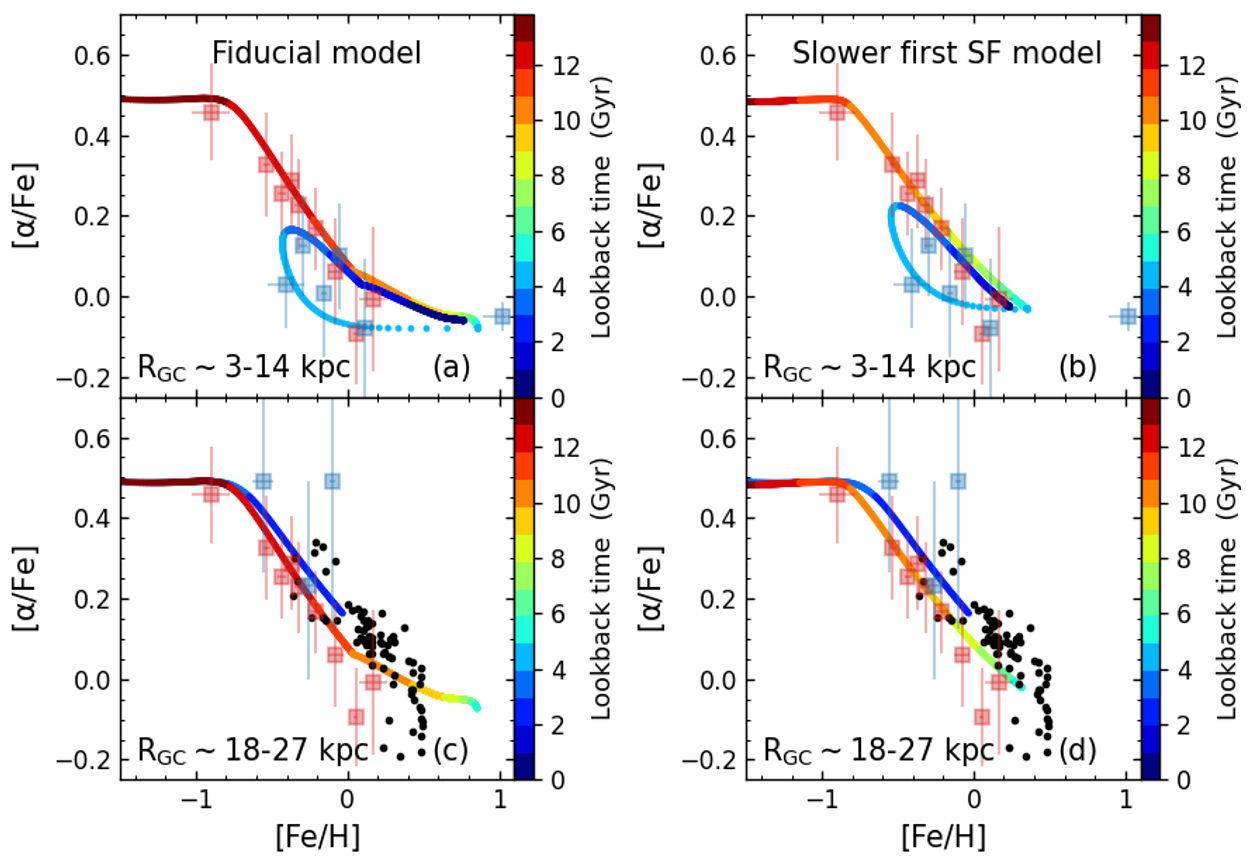}
    \caption{The [$\alpha$/Fe]--[Fe/H] diagrams for the M31 chemical evolution models coloured by lookback time (in Gyr)
    for $R_{\rm GC}\sim$3--14~kpc (upper panels) and for $R_{\rm GC}\sim$18--27~kpc (lower panels), comparing to observational results: the binned PN values computed from the data in \citet[][red and blue squares with errorbars respectively for $>4.5$ and $\sim2.5$ Gyr]{Bhattacharya22} using the models, and JWST's RGB stars from \citet[][black points]{nidever23} at $R_{\rm GC}\sim$18~kpc. The left panels are for the fiducial model, while the right models have a weaker initial star formation (see the main text for the details).
    }
    \label{fig:gce}
\end{figure*}

Figure \ref{fig:spat} shows the spatial distribution of our PN sample (red and blue points)  comparing to the JWST field from \citet[yellow square]{nidever23}.
The location of the JWST pointing is close to the region called ``the ring of fire'' in M31. This is the region ($R_{\rm GC}\sim$ 14--18 kpc) that shows a very strong star burst in the FUV map \citep{Kang09}.
There is an imaging survey, the Panchromatic Hubble Andromeda Treasury (PHAT, \citealt{dal12}, green open squares), with the Hubble Space Telescope (HST), which well covers the inner disk sample of PNe.
Ages and metallicities of the stellar populations are estimated \citep{williams17}, from which we constructed our chemical evolution models of M31 (\S 2.2).

In addition, a spectroscopic survey with the Dark Energy Spectroscopic Instrument (DESI, \citealt{dey23}, dashed yellow circles) is also marked, from which more accurate estimates of metallicities and possibly [$\alpha$/Fe] ratios may be available in future.
The Prime Focus Spectrograph (PFS) on the Subaru Telescope \citep{pfs} will also be used to map [$\alpha$/Fe] ratios in fa wide area of disk and halo of M31.

\subsection{GCE modelling of M31}

The GCE models we use in this paper were constructed in order to match the observed star formation history (SFH) and metallicity distribution function (MDF) from the PHAT survey \citep{williams17}, as well as the O/Ar ratios of PNe; the detailed model parameters were given in Table B.1 of \citet{Arnaboldi22}. Briefly, we used a GCE code \citep{kob00} but including the latest nucleosynthesis yields of asymptotic giant branch (AGB) stars, super-AGB stars, core-collapse supernovae \citep{kob20sr}, and SNe Ia \citep{kob20ia}, which allows us to accurately predict all elements including Ar and to convert O/Ar of PNe to [$\alpha$/Fe].
The standard initial mass function from \citet{kro08} is adopted.
The SN Ia model is taken from \citet{kob09}, which includes a metallicity effect in \citet{kob98}.

The SN Ia progenitors are C+O white dwarfs (WDs) from $\sim$3--8$M_\odot$ stars (at $Z=Z_\odot$), but the delay-times from the formation of the binary systems to the SN Ia explosions depend on binary physics. 
For the single-degenerate scenario, the delay-times are comparable to the lifetimes of the companion stars, while for the double-degenerate scenario, the delay-times can be as short as 34 Myrs (in the case where the secondary WD comes from a $8M_\odot$ star), or long for wide binaries.
Although delay-time distributions can be calculated with binary population synthesis models, there is no model that can reproduce the observations in the solar neighborhood \citep{kob22}. 
Triple and higher-order multiple systems could also affect the delay-time distributions.

Therefore, we used the analytic formula from \citet{kob09} that can reproduce the observations in the MW, as well as the delay-time distribution deduced from SN Ia rates in galaxies.
The adopted yields are for Chandrasekhar-mass explosions  from \citet{kob20ia}.
These result in more rapid decrease of the $\alpha$/Fe ratio than in other models such as \citet{chi97}. 
With our SN Ia model, the $\alpha$/Fe bimodality can be produced even without two distinct inflows in cosmological simulations \citep{kob23book}, although it would become clearer with two distinct inflows.
This could be tested with an IFU survey of edge-on disk galaxies such as the GECKOS survey \citep{GECKOS}.

\section{Results}

Figure \ref{fig:gce} shows the [$\alpha$/Fe]--[Fe/H] relations at $R_{\rm GC}\le14$ kpc (upper panels) and at $R_{\rm GC}$ $\ge$ 18 kpc (lower panels), comparing to the PN data converted from \citet{Bhattacharya22} and the JWST data from \citet{nidever23}.
Here we define [$\alpha$/Fe] $\equiv$ ([O/Fe]$+$[Mg/Fe]$+$[Si/Fe]$+$[S/Fe]$+$[Ca/Fe])$/5$ for the GCE model curves and the PN data (see next paragraph).
The fiducial model (left panels) is in excellent agreement with the observed SFH and MDF of the PHAT region, as well as the O/Ar--Ar/H relation of PNe.
The other model (right panels) gives a similar O/Ar--Ar/H relation, also explaining the lack of super-solar metallicity stars in the PHAT's MDF (see Fig.\,B1 of \citealt{Arnaboldi22}).

The observed O/Ar and Ar/H of PNe are converted into [$\alpha$/Fe] and [Fe/H], respectively, using the GCE models.
First, from each GCE model curve, we compute an affine transformation matrix to convert any point ($x$, $y$) in the $12+\log$ Ar/H vs $\log$ O/Ar plane to ($x'$, $y'$) in the [$\alpha$/Fe]--[Fe/H] plane.
We then apply it to the observed abundance ratios of individual PNe from \citet{Bhattacharya22}.
Finally, the obtained abundance ratios are binned as a function of [Fe/H], similar to the binning of $12+\log$ Ar/H with a velocity dispersion cut in \citet{Arnaboldi22}.
The error bars for [Fe/H] are the bin-width containing at least 15 thick-disk PNe (or 8 thin-disk PNe); the error for [$\alpha$/Fe] is the transformed measurement error added to standard deviation in quadrature.

Our M31 models assume two distinct star formation episodes triggered by two gas infalls, similar to those in \citet{chi97}.
The first star formation is probably caused by cosmological accretion with [Fe/H] $<-2.5$ in order to explain the PHAT's MDF.
As a result, the predicted [$\alpha$/Fe] ratio shows a plateau with a constant value of $\sim 0.5$, and sharply decreases from [Fe/H] $\sim -0.8$, in contrast to the [$\alpha$/Fe] `knee' at [Fe/H] $\sim -1$ in the solar neighborhood \citep[\S 1; see also][]{kob20sr}.
This means that M31 has an early star formation with a higher star formation rate than in the MW, possibly because of M31's twice larger total mass (\S 2.1).
In our fiducial model, the metallicity reaches [Fe/H] $\sim+0.8$ (left panels) while only $\sim+0.4$ with relatively weaker initial star formation (right panels).

As in the observed SFH, the secondary gas infall is set at 9 Gyr after the onset of the star formation, i.e., about 4.5 Gyr ago, which lasts until about 2 Gyr ago.
This is probably caused by a wet merger of a fairly large gas-rich satellite galaxy.
This causes a `loop' in the [$\alpha$/Fe]--[Fe/H] diagram (upper panels):
(1) Dilution due to metal-poorer gas lowers the metallicity, (2) star burst quickly increases the [$\alpha$/Fe] ratio, and then (3) SNe Ia bring the track back to the original.
Late infall of metal poorer gas was also shown by the resolved stellar population studies of the outer disks of M31  \citep{Bernard12,Bernard15}.

The low-$\alpha$ (and low O/Ar) component is seen in our PN sample only at $R_{\rm GC}\le14$ kpc. For the outer disk, we provided another model assuming a secondary star burst about 4.5 Gyr ago completely from a primordial gas inflow (lower panels), in order to explain the high O/Ar ratios of our thin-disk PNe at $R_{\rm GC}\ge18$ kpc (Fig.\,6 of \citealt{Arnaboldi22}).
In the lower panels, we show the [$\alpha$/Fe]--[Fe/H] relations predicted from the outer disk models.
After the same evolutionary track as in the upper panels, the secondary star formation forms young stars with high $\alpha$/Fe (and high O/Ar) ratios.
This gas accretion could be associated with the wet merger event mentioned above, but the metallicity might be kept very low due to a smaller amount of pre-enriched gas in the outer disk.
The young high-$\alpha$ population gives only slightly higher $\alpha$/Fe than the old high-$\alpha$ population,
and is not as distinct as in the $\alpha$/Fe bimodality in the solar neighborhood of the MW.
Instead, we predict a slight upward shift in [$\alpha$/Fe] ratios for a younger population than 4.5 Gyr.
Unlike PNe, RGB stars are insensitive to the young population \citep{dorman15}, and the JWST observation is in fact consistent with the first, single star burst at $\gtsim4.5$ Gyr.

It is important to note that the high O/Ar population of PNe suggests that the old high-$\alpha$ population exists at all radii at $R_{\rm GC}<30$ kpc.
Similar to O/Ar, the $\alpha$/Fe ratio decreases due to the delayed enrichment from SNe Ia.
Below the [$\alpha$/Fe] slope, the young low-$\alpha$ population is found only in the inner disk ($R_{\rm GC}\le14$ kpc).
On the other hand, the outer disk seems to contain {\it young} high-$\alpha$ population, slightly {\it above} the [$\alpha$/Fe] slope.
The location of the young population in the [$\alpha$/Fe]--[Fe/H] diagram varies as a function of the radial ranges.

\begin{figure}
    \includegraphics[width=\columnwidth]{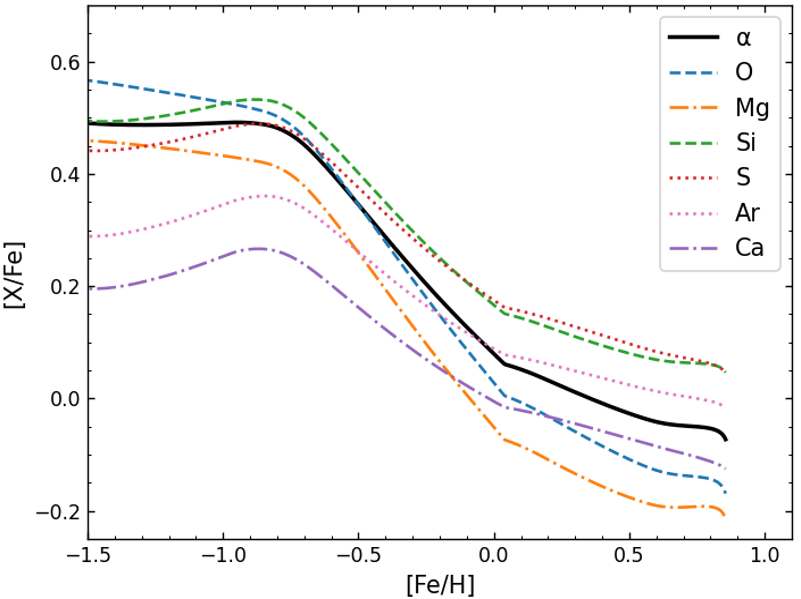}
    \caption{The evolution of elemental abundances [X/Fe] for O, Mg, Si, S, Ar and Ca (as well as  [$\alpha$/Fe] $\equiv$ ([O/Fe]$+$[Mg/Fe]$+$[Si/Fe]$+$[S/Fe]$+$[Ca/Fe])$/5$) against [Fe/H] for the fiducial model of M31's old thicker-disk (lookback time $>$ 4.5~Gyr) as shown in Figure~\ref{fig:gce} (c).
    }
    \label{fig:alpha}
\end{figure}

Our $\alpha$/Fe plateau value seems to be consistent with \citet{escala20}, who showed 18 RGB stars at $R_{\rm GC}=31$ kpc spanning [$\alpha$/Fe] from 0 to $+1$ around [Fe/H] $=-1$.
Note that, however, the detailed comparison should be made not with $\alpha$/Fe but with elemental abundances.
The definition of [$\alpha$/Fe] was not given in \citet{nidever23}, but it can include Mg, Si, maybe O and Ca, not much S, according to the available lines in the spectral coverage.
Figure \ref{fig:alpha} compared the [$\alpha$/Fe] tracks in our fiducial model of M31's old thicker-disk for various $\alpha$ elements.
O gives the steepest slope, while Ca gives a shallow slope, mainly due to a different contribution from SNe Ia \citep{kob20sr,kob20ia}.
The [$\alpha$/Fe] ratio could have a systematic offset of $0.3$ dex depending on the exact definition.
For RGB stars, more detailed analysis with higher resolution of spectra would be needed to constrain the history of M31.

\section{Conclusions}

We predict an $\alpha$/Fe bimodality in M31 using our state-of-the-art galactic chemical evolution models that allow us to compare the PN observation (with emission lines) to that of RGB stars by JWST. As well as these elemental abundances, the models also well explain the stellar populations in the inner disk estimated with the HST (see Appendix B of \citealt{Arnaboldi22}).

M31 had a less quiet merger history than the MW, as discussed in previous works \citep[][and references therein]{yin09,Bhattacharya23a}.
In this Letter, we find that the high-$\alpha$ thicker-disk population at all radii in $R_{\rm GC} < 30$ kpc, formed by an initial star burst more intense than the MW.
This results in the so-called $\alpha$/Fe `knee' appears at a higher metallicity [Fe/H] $\sim -0.8$, in contrast to $\sim-1$ in the solar neighborhood of the MW.
Then, M31 underwent a secondary star formation about 2--4.5 Gyr ago, probably triggered by a wet merger of a fairly large gas-rich satellite
galaxy. The dilution causes a `loop' in the [$\alpha$/Fe]--[Fe/H] diagram.
This young low-$\alpha$ thin-disk is seen only in the inner disk ($R_{\rm GC} < 14$ kpc).

The MW thin-disk is made of the low-$\alpha$ population in the solar neighborhood, which becomes dominant at outer regions.
In the outer disk of M31 ($R_{\rm GC} > 18$ kpc), however, our PN observation suggests that accretion of almost pristine gas started a new chemical evolution track, making a slightly higher-$\alpha$ young ($\sim2.5$ Gyr) population at a given metallicity; the JWST observation at $R_{\rm GC} \sim 18$ kpc would not detect this $\alpha$/Fe bimodality in M31 because of its observed location and of the age bias for RGB stars.

The appearance of the $\alpha$/Fe bimodality depends on the merging history at various galactocentric radii.
In order to reveal the full story of M31, it is important to use elemental abundances not only of RGB stars but also of PNe covering a wide range of ages, with wide-field multi-object spectroscopy such as DESI and the Subaru PFS.

\begin{acknowledgments}
We would like to thank E. Kirby and the Subaru's PFS Galactic Archaeology survey team for fruitful discussion.
CK acknowledges funding from the UK Science and Technology Facility Council through grant ST/R000905/1, ST/V000632/1. 
The work was also funded by a Leverhulme Trust Research Project Grant on ``Birth of Elements''.
SB is funded by the INSPIRE Faculty award (DST/INSPIRE/04/2020/002224), Department of Science and Technology (DST), Government of India.
SB, MAR and OG would like to acknowledge support to this research from the European Souther Observatory, Garching, through the 2021 SSDF and the Excellence Cluster ORIGINS, which is funded by the Deutsche Forschungsgemenschaft (DGF, German Research Foundation) under Germany's Excellence Strategy - EXC-2094-390783311.
\end{acknowledgments}

\bibliography{ms}{}
\bibliographystyle{aasjournal}

\end{document}